\documentclass[twocolumn,showpacs,preprintnumbers,amsmath,amssymb,superscriptaddress]{revtex4}
\usepackage[bookmarks, pdftitle={ms_spinel_v12}, pdfauthor={Nair}, colorlinks=true, linkcolor=black, citecolor=blue, urlcolor=black]{hyperref}
\usepackage{amssymb}
\usepackage{amsmath}
\usepackage{graphicx}
\usepackage{dcolumn}
\usepackage{bm}
\usepackage{times}
\begin{document}
\preprint{APS/123-QED}
\title
{Spin Freezing in the Spin Liquid Compound FeAl$_2$O$_4$}
\author{Harikrishnan S. Nair}
\email{h.nair.kris@gmail.com, hsnair@uj.ac.za}
\affiliation{Highly Correlated Matter Research Group, Physics Department, P. O. Box 524, University of Johannesburg, Auckland Park 2006, South Africa}
\author{Ramesh Kumar K.}
\affiliation{Highly Correlated Matter Research Group, Physics Department, P. O. Box 524, University of Johannesburg, Auckland Park 2006, South Africa}
\author{Andr\'{e} M. Strydom}
\affiliation{Highly Correlated Matter Research Group, Physics Department, P. O. Box 524, University of Johannesburg, Auckland Park 2006, South Africa}
\affiliation{Max Planck Institute for Chemical Physics of Solids (MPICPfS), N\"{o}thnitzerstra{\ss}e 40, 01187 Dresden, Germany}
\date{\today}
\begin{abstract}
Spin freezing in the $A$-site spinel FeAl$_2$O$_4$ which is a spin liquid candidate is studied using remnant magnetization and nonlinear magnetic susceptibility and isofield cooling and heating protocols. The remnant magnetization behavior of FeAl$_2$O$_4$ differs significantly from that of a canonical spin glass which is also supported by analysis of the nonlinear magnetic susceptibility term $\chi_3 (T)$. Through the power-law analysis of $\chi_3 (T)$, a spin-freezing temperature, $T_g$ = 11.4$\pm$0.9~K and critical exponent, $\gamma$ = 1.48$\pm$0.59 are obtained. Cole-Cole analysis of magnetic susceptibility shows the presence of broad spin relaxation times in FeAl$_2$O$_4$, however, the irreversible dc susceptibility plot discourages an interpretation based on conventional spin glass features. The magnetization measured using the cooling-and-heating-in-unequal-fields protocol brings more insight to the magnetic nature of this frustrated magnet and reveals unconventional glassy behaviour. Combining our results, we arrive at the conclusion that the present sample of FeAl$_2$O$_4$ consists of a majority spin liquid phase with "glassy" regions embedded. 
\end{abstract}
\maketitle
%
\indent
$AB_2X_4$ ($A$ = Mn, Fe, Co; $B$ = Al, Sc, Rh; $X$ = O, S) compounds where a magnetic atom occupies the tetrahedrally coordinated $A$ site are known as $A$-site spinels and 
are frustrated magnets where the frustration effects arise from competing nearest neighbor ($n.n.$) and next-nearest neighbor ($n.n.n.$) exchange 
interactions\cite{tristanprb_72_174404_2005geometric,krimmelprb_79_134406_2009spin, fichtl_prl_94_027601_2005,fritsch_prl_92_116401}. The spinel structure is 
composed of diamond lattice formed by the $A$ site atoms and pyrochlore network of the $B$ site. The ideal diamond lattice with only $n.n$ interactions is not geometrically 
frustrated in contrast with the pyrochlore lattice which is inherently geometrically frustrated. Additional $n.n.n.$ interactions are necessary to create frustration in the diamond 
lattice. In $A$-site spinels, frustration in the diamond lattice can manifest as "spiral" spin liquid\cite{bergmannature_3_487_2007order}, spin 
liquid\cite{krimmelprb_79_134406_2009spin,zaharko_2011spin}, orbital liquid\cite{fritsch_prl_92_116401}, orbital glass\cite{fichtl_prl_94_027601_2005} or a spin-orbital 
singlet state with quantum critical point\cite{chen_prl_102_096406_2009spin} and hence these materials are of immense interest to condensed matter physicists.
Specifically, the theoretical work by Bergmann {\it et al.,}\cite{bergmannature_3_487_2007order} which invokes the mechanism of "order by disorder"\cite{villain_jp_41_1263_1980order} as the degeneracy-breaking mechanism, predicts the emergence of a "spiral" spin liquid in the $A$-site spinels which is characterized by a manifold of degenerate ground states in a system devoid of defects. In a subsequent work, the influence of quenched random impurities such as a random bond, a vacancy or an interstitial spin on the "spiral" spin liquid properties was undertaken\cite{savary_prb_84_064438_2011impurity}. It was found that quenched disorder can act as a degeneracy-breaking mechanism. A "swiss cheese model" was introduced which explained, to some extent, the contrasting findings of long-range ordered and "glassy" magnetic ground states reported for the spinel CoAl$_2$O$_4 $\cite{tristanprb_72_174404_2005geometric,macdougall_pnas_108_15693_2011kinetically}. The fact that the magnetic properties of $A$-site spinels are governed by two factors -- frustration and site disorder -- has been reiterated through experimental studies on Al-based systems\cite{roy_prb_88_174415_2013experimental,nair,hanashima_jpsj_82_024702_2013spin}. A magnetic phase diagram for CoAl$_2$O$_4$ tuned by the defect content ($\eta$, the inversion parameter) has been proposed, where spin liquid and spin glass phases compete as a function of $\eta$\cite{hanashima_jpsj_82_024702_2013spin}. 
\\
\begin{figure}[!t]
\centering
\includegraphics[scale=0.65]{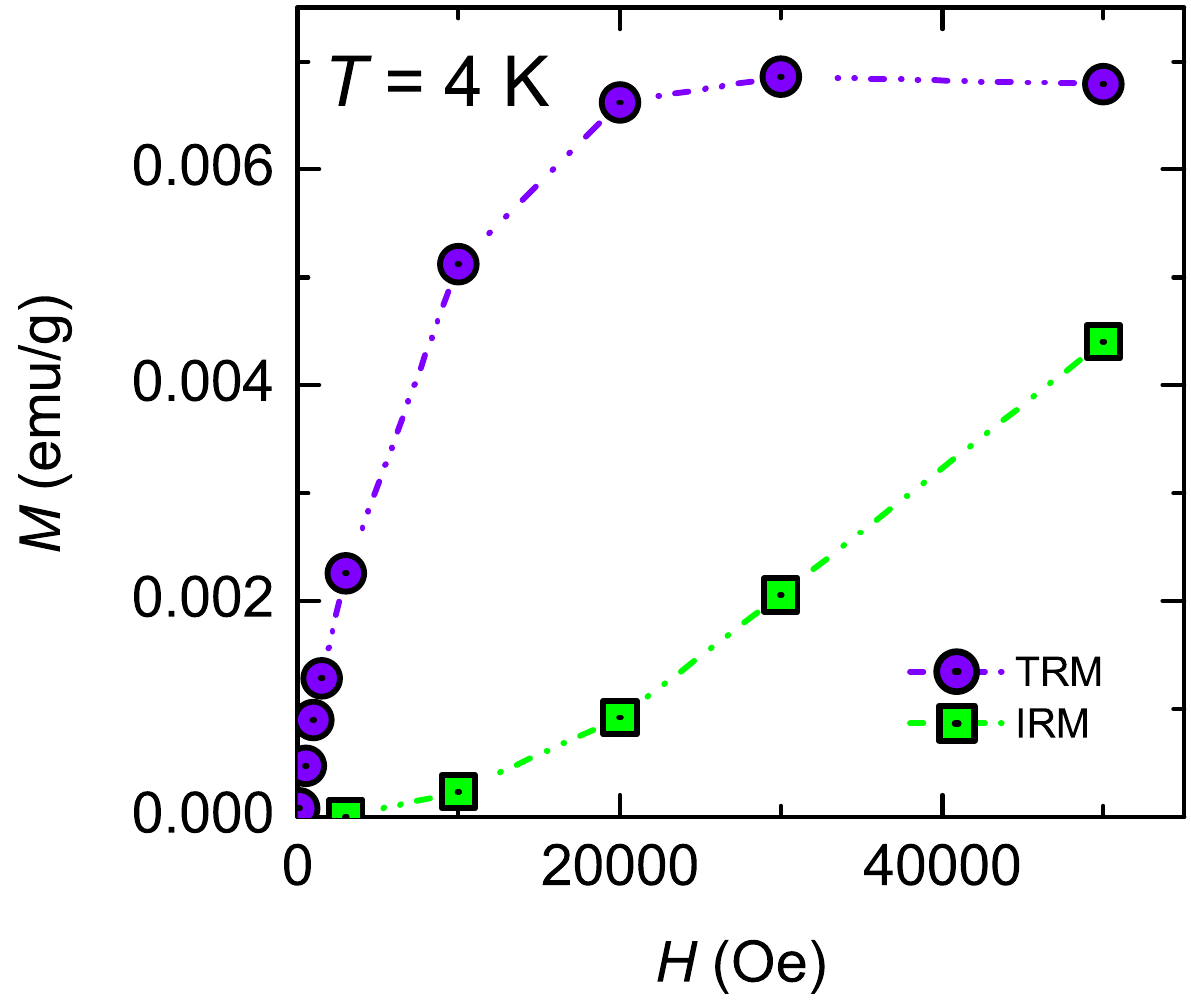}
\caption{\label{fig_trm} (color online) The thermoremanent (TRM) and the isothermoremanent (IRM) curves for FeAl$_2$O$_4$ at 4~K as a function of applied magnetic field. The TRM and IRM responses of FeAl$_2$O$_4$ are characteristically different from that of a canonical spin glass or a superparamagnet (see \cite{benitez_prb_83_134424_2011fingerprinting} for a comparison of TRM and IRM curves of different magnetic systems).}
\end{figure}
%
%
%
\begin{figure}[!t]
\centering
\includegraphics[scale=0.55]{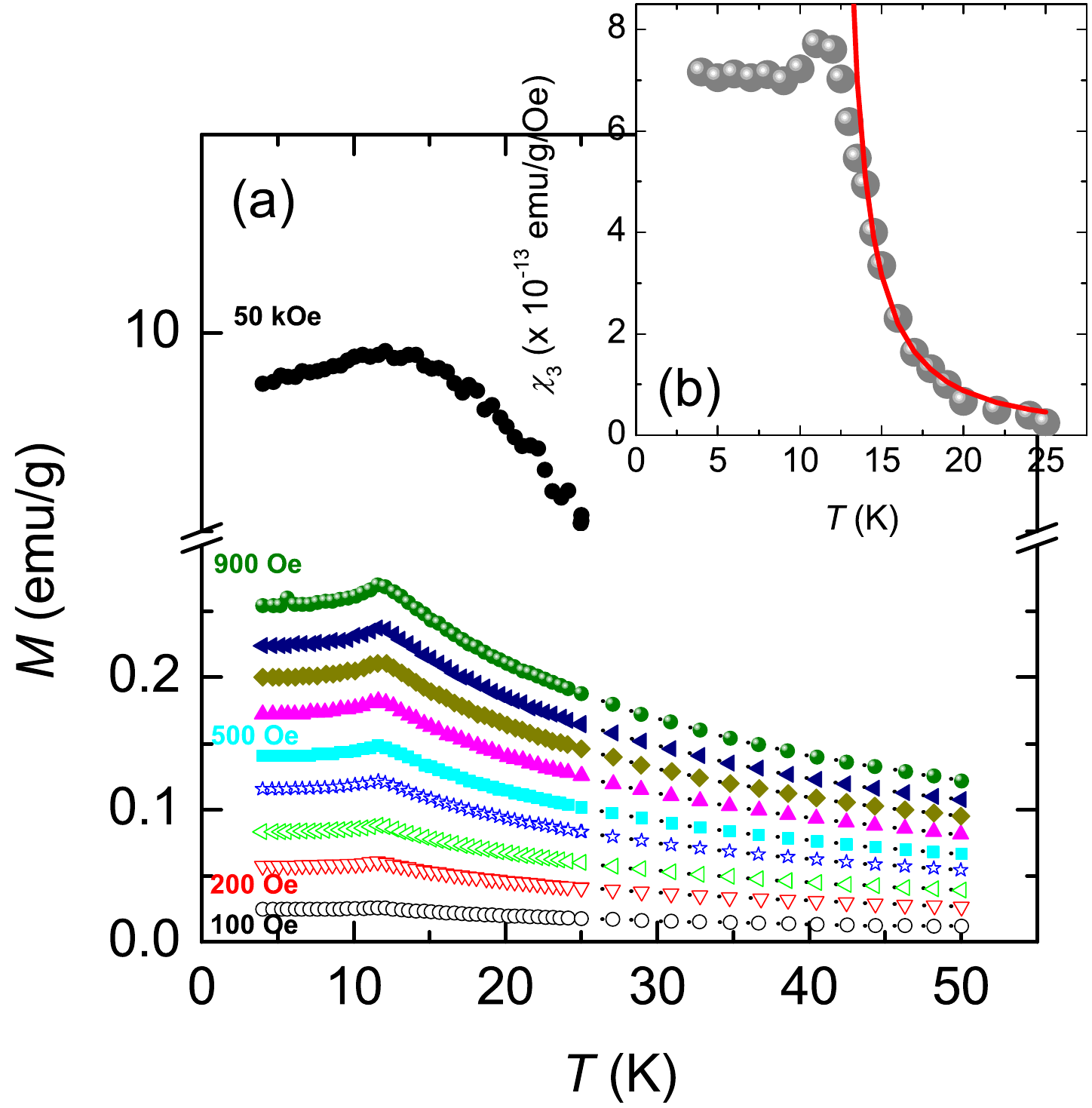}
\caption{\label{fig_fc} (color online) (a) The temperature dependence of field-cooled dc magnetization plots of FeAl$_2$O$_4$ at various applied fields in the range 100~Oe (black open circles, bottom curve) to 50~kOe (black closed circles, top curve). $M$ versus $H$ plots at different temperature values are extracted from this data. (b) Shows the third harmonic of nonlinear susceptibility, $\chi_3$, along with a power-law fit (solid line) using $\chi_3$ = $\chi^0_3 \left(T/T_g - 1\right)^{-\gamma}$.}
\end{figure}
%
\indent
The effect of weak disorder on geometrically frustrated lattices has been treated theoretically in the parlance of fully-frustrated 
Ising\cite{campbell_prb_52_R9819_1995frustration} and Heisenberg systems\cite{saunders_prl_98_157201_2007spin,andreanov_prb_81_014406_2010}. Though these 
theoretical studies have pointed towards the realization of a spin glass state at low temperature due to the weak disorder effect, the question whether additional random 
interactions on the top of frustration would lead to a "true" spin glass phase at low temperatures remains relevant\cite{villain_zp_33_31_1979insulating}. Our previous studies on 
the $A$-site spinels FeAl$_2$O$_4$ and MnAl$_2$O$_4$ using magnetometry and polarized neutron diffraction have revealed the presence of significant spin correlations 
arising purely from frustration effects\cite{nair}. MnAl$_2$O$_4$ was observed to posses an ordered magnetic state below $\approx$ 45~K whereas FeAl$_2$O$_4$ exhibited only 
short-range order until down to 4~K. In the present work, we study the spin freezing in FeAl$_2$O$_4$ through ac and dc nonlinear susceptibility and 
thermoremanent (TRM) and isothermoremanent (IRM) magnetization along with other magnetization protocols with the aim of elucidating the magnetometric "signature" of a spin liquid. To the best of our knowledge, it is a first attempt to use the TRM/IRM signatures to understand a spin liquid system.
\\
\indent
The magnetic measurements reported in this work were all performed on polycrystalline pellets (the details of the sample-preparation are discussed in our previous publication [\onlinecite{nair}]) using a commercial SQUID magnetometer, Quantum Design Inc. In order to 
measure TRM, the sample was field-cooled down to 4~K from 300~K; the magnetic field was switched off and the remnant magnetization was recorded instantly. Similarly, for 
IRM, the sample was cooled in zero field down to 4~K from 300~K, the magnetic field was switched on and instantly switched off to measure the remnant 
magnetization\cite{benitez_prb_83_134424_2011fingerprinting}. The measurements were repeated for different values of applied magnetic field. In Fig~\ref{fig_trm} the 
field dependence of TRM/IRM of FeAl$_2$O$_4$ are presented. The remnant magnetization curves are evidently distinguished from those of canonical spin glasses but resemble that of a diluted antiferromagnet in field (DAFF)\cite{benitez_prb_83_134424_2011fingerprinting} albeit with some notable differences. In the case of DAFF, the IRM response is expected to be identically zero for all fields and the TRM is normally inversely proportional to the domain size. In FeAl$_2$O$_4$ however, IRM-response  is not zero but reaches up to a value which is half that of the TRM-response at high fields. The field-stability observed in the case of FeAl$_2$O$_4$ is a notable feature especially when the remnant magnetization response does not show any similarity to canonical spin glasses\cite{benitez_prb_83_134424_2011fingerprinting}.
\\
\indent 
In order to understand more about the spin-freezing process in FeAl$_2$O$_4$, we performed dc magnetic measurements to extract the nonlinear susceptibility terms. The nonlinear terms of dc susceptibility are sensitive to the spin freezing order parameter. In order to estimate the nonlinear contributions, the magnetization data were obtained under field-cooled conditions in the range 3--50~K for different values of applied magnetic fields in the range 100~Oe - 50~kOe, see Fig~\ref{fig_fc}. Prior to each measurement, the sample was heated up to 150~K and then field-cooled to lowest temperature in order to measure the magnetization as a function of temperature. The nonlinear susceptibilities are extracted from the magnetization data by writing:
\begin{eqnarray}
M/H(T) = \chi_1(T) - \chi_3(T)H^2 + \mathcal{O}(H^4)\\
= \chi_1 (T) - a_3(T)\chi^3_1 (T)H^2 + \mathcal{O}(H^4)\\
\chi_{nl}(T,H) = 1 - M(T, H)/\chi_1H
\label{eqn_nonlin}
\end{eqnarray}
where $\chi_1(T)$ is the linear susceptibility at temperature $T$, $\chi_3(T)$ is the third harmonic of nonlinear susceptibility, the coefficient $a_3$ = $\chi_3/(\chi_1)^3$ 
and $\chi_{nl}$ is the net nonlinear susceptibility\cite{gingras_prb_78_947_1997static}. The nonlinear terms of susceptibility were extracted from the data following a polynomial fit to the magnetization using $M = \chi_1 H - \chi_3 H^3 + \chi_5 H^5$. The $\chi_1$ and $\chi_3$ terms extracted from the fit (see the inset of Fig~\ref{fig_fc}) show a peak at $T_a \approx$ 13~K ($T_a$ is designated as the temperature at which a peak is observed in the magnetic response in Fig~\ref{fig_fc}). In order to test static criticality wherein the $\chi_3$ term should diverge, a fit was administered using the power-law, $\chi_3$ = $\chi^0_3 \left(T/T_g - 1\right)^{-\gamma}$. 
%
\begin{figure}[!b]
\centering
\includegraphics[scale=0.45]{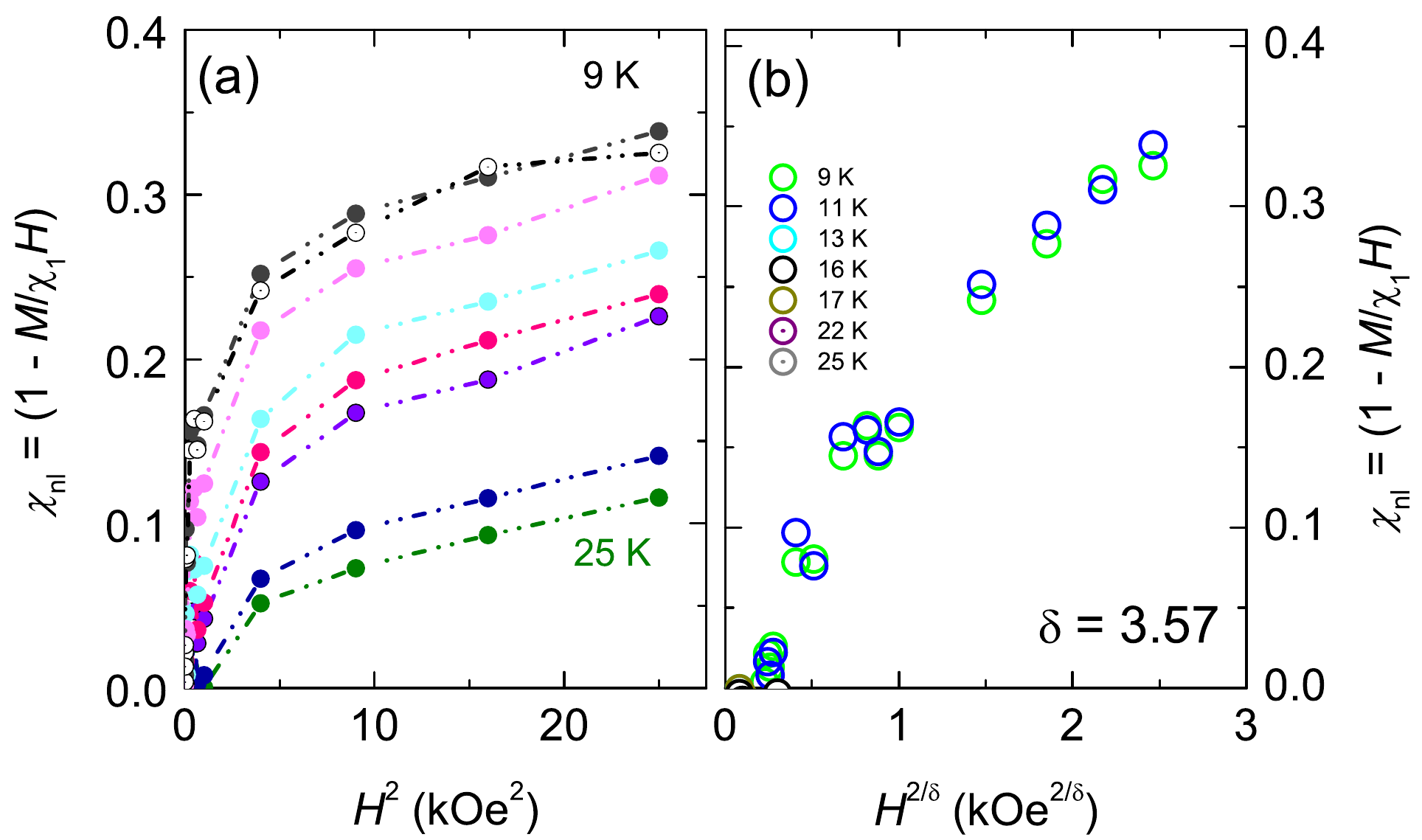}
\caption{\label{fig_nonlin} (color online) (a) The nonlinear susceptibilities, $\chi_{nl}$, as a function of $H^2$ at different selected temperatures above and below $T_a$ (which is identified as the peak observed in dc magnetization curves in Fig~\ref{fig_fc}). (b) The scaling plot of the data presented in (a) with $\delta$ = 3.57 clearly brings out the data collapse conforming to the scaling exponent, $\delta$ (see text).}
\end{figure}
%
Here, $T_g$ is the glass-transition temperature and $\gamma$ is a critical exponent. The fit yielded $T_g$ = 11.4$\pm$0.9~K, $\gamma$ = 1.48$\pm$0.59.  The $\gamma$ value is smaller than that observed generally in spin glasses\cite{jonsson_prl_75_4138_1995aging} but much greater than that for a mean-field system. The net nonlinear susceptibility was obtained as $\chi_{nl} = \left(1 - M/\chi_1H \right)$. A plot of $\chi_{nl}$ versus $H^2$ is presented in Fig~\ref{fig_nonlin}. From the power law dependence, $\chi_{nl}\left(T = T_g, H\right) \propto H^{2/\delta}$, an estimate of the critical exponent $\delta$ can be obtained as the slope of the plot of natural logarithm of $\chi_{nl}$ and $H$. This led to an estimate of $\delta$ = 3.57$\pm$03 in the present case. The low-field limiting case, when writing $\chi_{nl}$ = $H^{2/\delta} f (\tau^{(\gamma + \beta)/2}/H)$, where $\tau$ = ($T/T_g$ - 1), is $\chi_{nl}$ = $H^{2/\delta}$. $\tau^{-2\gamma/(\gamma + \beta)}$, then $\delta$ = 1 + $\gamma$/$\beta$. In Fig~\ref{fig_nonlin} (b), a scaling plot with $\delta$ = 3.57 demonstrates the data collapse onto a universal plot for temperatures below 25~K.
\\
\indent
From the neutron diffraction studies using polarized neutrons\cite{nair}, it is clear that short-range spin-spin correlations assume importance in this material. The short-range magnetic order in frustrated magnets reflect as a statistical distribution of relaxation times where each cluster acts as independent unit\cite{huser_jpc_19_3697_1986phenomenological}. This distribution of relaxation times can be experimentally extracted through the analysis of complex ac magnetic susceptibilities. For this purpose, the real and imaginary parts of the complex ac magnetic susceptibilities were recorded for FeAl$_2$O$_4$ using a commercial Physical Property 
Measurement System. The ac susceptibility data, $\chi (f, T)$, did not present a clear shift in the peak-value with frequency (presented in \cite{nair}) as is normally observed in the case of canonical spin glasses\cite{mydosh_book}. However, in the presence of applied dc magnetic field, the peak in $\chi (f, T)$ was observed to broaden. At this point it is interesting to note the case of another frustrated magnet Dy$_2$Ti$_2$O$_7$, which is a spin ice, where the application of dc magnetic field raised the spin freezing temperature\cite{snyder_nature_413_48_2001spin}. The spin dynamics in Dy$_2$Ti$_2$O$_7$ spin ice in the limit of very low disorder was peculiar with very narrow range of relaxation times and was claimed to represent a new form of spin freezing in a frustrated magnet.
\\
%
%
\begin{figure}[!t]
\centering
\includegraphics[scale=0.30]{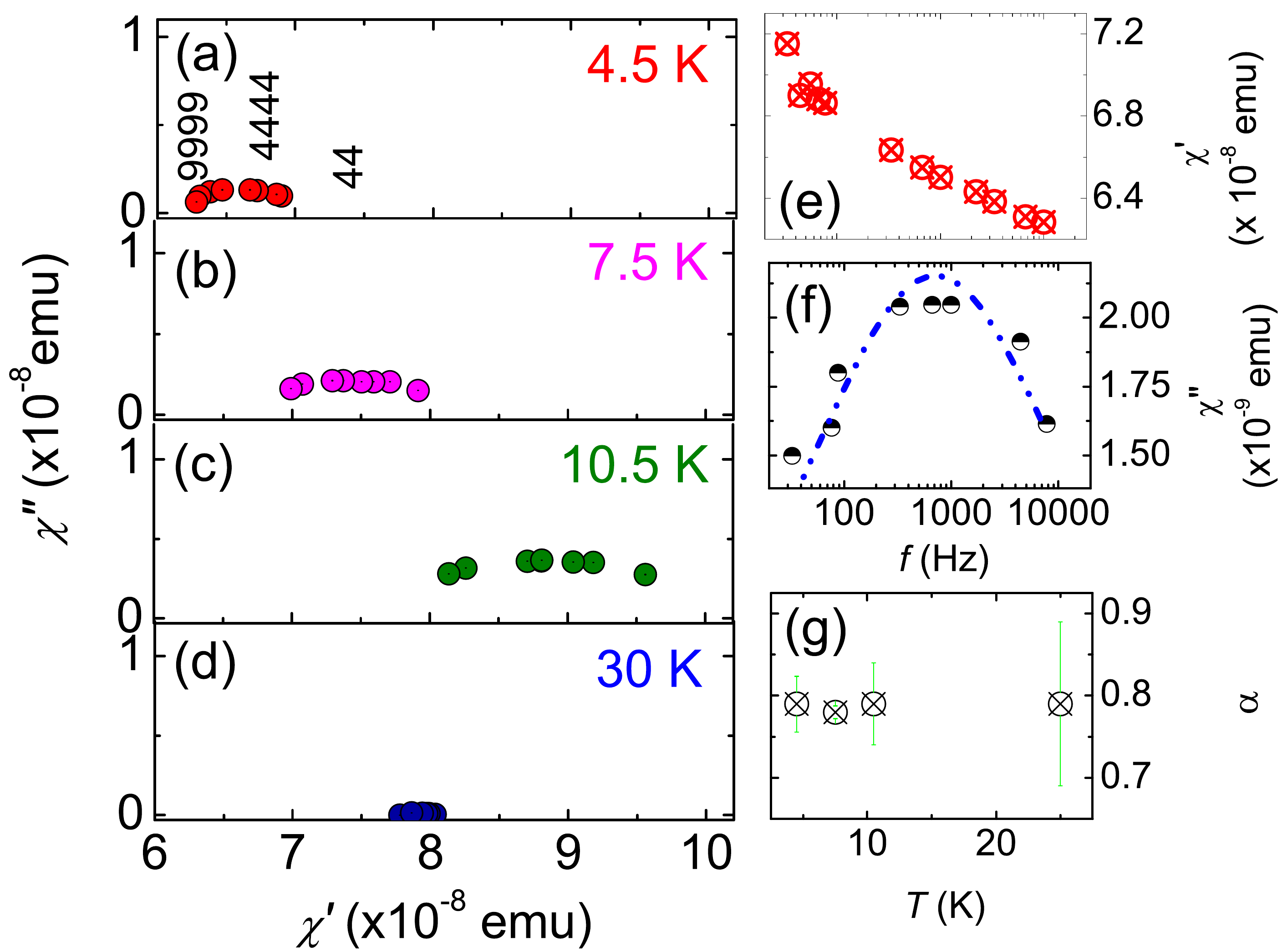}
\caption{\label{fig_cole_a} (color online) (a-d) The Cole-Cole plot ($\chi"$-$\chi'$) at different temperatures below 30~K supports the claim of a broad distribution of 
relaxation times in FeAl$_2$O$_4$. The numbers in (a) indicate the applied frequency in Hz.  The x-axis is offset from zero for clarity of the graph. (e) presents the dependence of 
$\chi'$ on frequency. (f) The imaginary part of the magnetic susceptibility, $\chi"$, as a function of frequency along with the theoretical curve according to Eqn~\ref{eqn_chi}. (g) 
The parameter $\alpha$ has a magnitude around 0.8 which indicates a broad relaxation.}
\end{figure}
%
\indent
The spin dynamics and the relaxation time distribution in frustrated or glassy magnets can be analysed using the Cole-Cole formalism which is the magnetic analogue of the Debye model developed for frequency dispersion of dielectric response\cite{cole_jcp_9_341_1941dispersion}. The relaxation of a magnetic system can be formulated as, 
\begin{equation}
\chi = \chi_s + [(\chi_0 - \chi_s)/(1 + (i\omega \tau_c)^{1 - \alpha})] 
\label{eq_cole}
\end{equation}
where $\chi_0$ and $\chi_s$ are the isothermal ($\omega$ = 0) and adiabatic ($\omega \rightarrow \infty$) susceptibilities respectively and $\tau_c$ is the median 
relaxation time around which the distribution of relaxation times occur. The parameter $\alpha$ describes the "flatness" of the distribution times; $\alpha$ = 1 corresponds to infinitely wide distribution whereas $\alpha$ = 0 gives the familiar Debye relaxation with a single relaxation time. The ideal case of single relaxation time can be contrasted from the case where a distribution of relaxation times are present\cite{dekker_prb_40_11243_1989activated}. Equation~(\ref{eq_cole}) can be decomposed to obtain the relation:
%
\begin{equation}
\chi" (\omega) = \chi^0
\left( 1 - \frac{\mathrm{sin}[(1/2)\beta\pi]}{\mathrm{cosh}[\beta \mathrm{ln}(\omega\tau_c)] + \mathrm{cos}[(1/2)\beta\pi]}\right)
\label{eqn_chi}
\end{equation}
where $\omega$ = 2$\pi f$ and $\chi^0 = \frac{\chi_0 - \chi_s}{2}$ and $\beta$ = (1 - $\alpha$).
The Cole-Cole plots obtained for FeAl$_2$O$_4$ are presented in Fig~\ref{fig_cole_a} (a-d) for the temperatures 4.5, 7.5, 10.5 and 30~K showing a "flat" spin relaxation compared to the case of Dy$_2$Ti$_2$O$_7$ where a semi-circular arc was obtained\cite{snyder_nature_413_48_2001spin}. A flat Cole-Cole plot for FeAl$_2$O$_4$ indicates broad relaxation times are present in the system. Fig~\ref{fig_cole_a} (e) shows the frequency variation of the real part of susceptibility. The imaginary part of susceptibility, $\chi" (f)$, as a function of frequency was fitted to the Eqn~(\ref{eqn_chi}) at various temperatures as shown in Fig~\ref{fig_cole_a} (f). The $\chi"$ does not follow a single relaxation time as per the Casimir-du Pr\'{e} relation $\chi"(f) = f \tau[(\chi_0 - \chi_s)/(1 + f^2 \tau^2)]$ used for spin ice\cite{snyder_nature_413_48_2001spin}. The parameter $\alpha$ extracted from the fit (Fig~\ref{fig_cole_a} (g)) displays a value around 0.79$\pm$0.05 at low temperatures thereby indicating the strong relaxation present, compared to the spin ice Dy$_2$Ti$_2$O$_7$ which had $\alpha \approx$ 0.5. Alternatively, the value of $\beta \approx$ 0.2 also implies a deviation from mean-field value.  
\\
%
%
\begin{figure}[!b]
\centering
\includegraphics[scale=0.55]{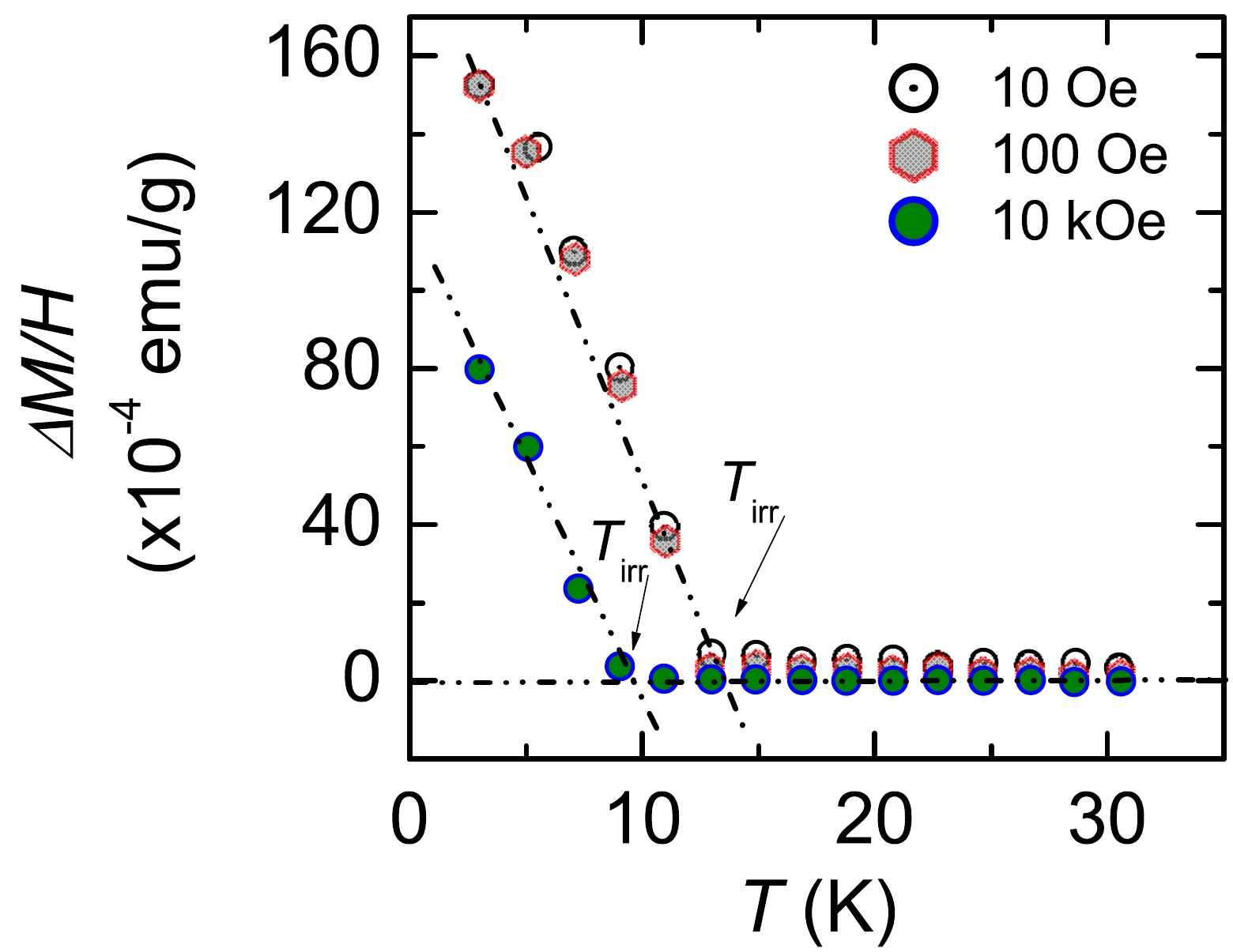}
\caption{\label{fig_dc} (color online) The "irreversible" dc susceptibility $\Delta M/H$ of FeAl$_2$O$_4$  where $\Delta M$ = ($M_{FC}$ -- $M_{ZFC}$). The 
irreversibility temperature is shown as $T_{irr}$ and is indicated by arrows. The dash-dotted lines are guides to the eye.}
\end{figure}
%
\indent
The importance of quenched disorder on the spin liquid properties of $A$-site spinels have been theoretically treated in the recent literature\cite{savary_prb_84_064438_2011impurity}. A systematic dependence of the ground state spin liquid properties of the spinel CoAl$_2$O$_4$ on the content of disorder, $\eta$, has been experimentally investigated recently\cite{hanashima_jpsj_82_024702_2013spin}. The proposed $T$--$\eta$ phase diagram shows that the spin liquid phase is stable only for low content of disorder. With the $\eta >$ 0.08, though high degeneracy of ground states exists in the spin liquid state, a spin glass ground state is selected by the system. 
The magnetic ordering in the spin glass phase in the presence of applied field can be studied by plotting the irreversible susceptibility, $\Delta M/H$ = ($M_{FC}$ -- $M_{ZFC}$)/$H$ as presented in Fig~\ref{fig_dc}. The disordered spin glass compositions of CoAl$_2$O$_4$ showed the occurrence of a "weak" and a "strong" irreversibility as slope-changes in $\Delta M/H$\cite{hanashima_jpsj_82_024702_2013spin}. Such two-step irreversibilities are commonly observed for canonical spin glasses, for example, CuMn\cite{kenning_prl_66_2923_1991irreversibility} and Y$_2$Mo$_2$O$_7$\cite{miyoshi_jpsj_69_3517_2000successive}. Any indication of such two-step irreversibilities are absent in Fig~\ref{fig_dc}.
\\
%
%
\begin{figure}[!t]
\centering
\includegraphics[scale=0.36]{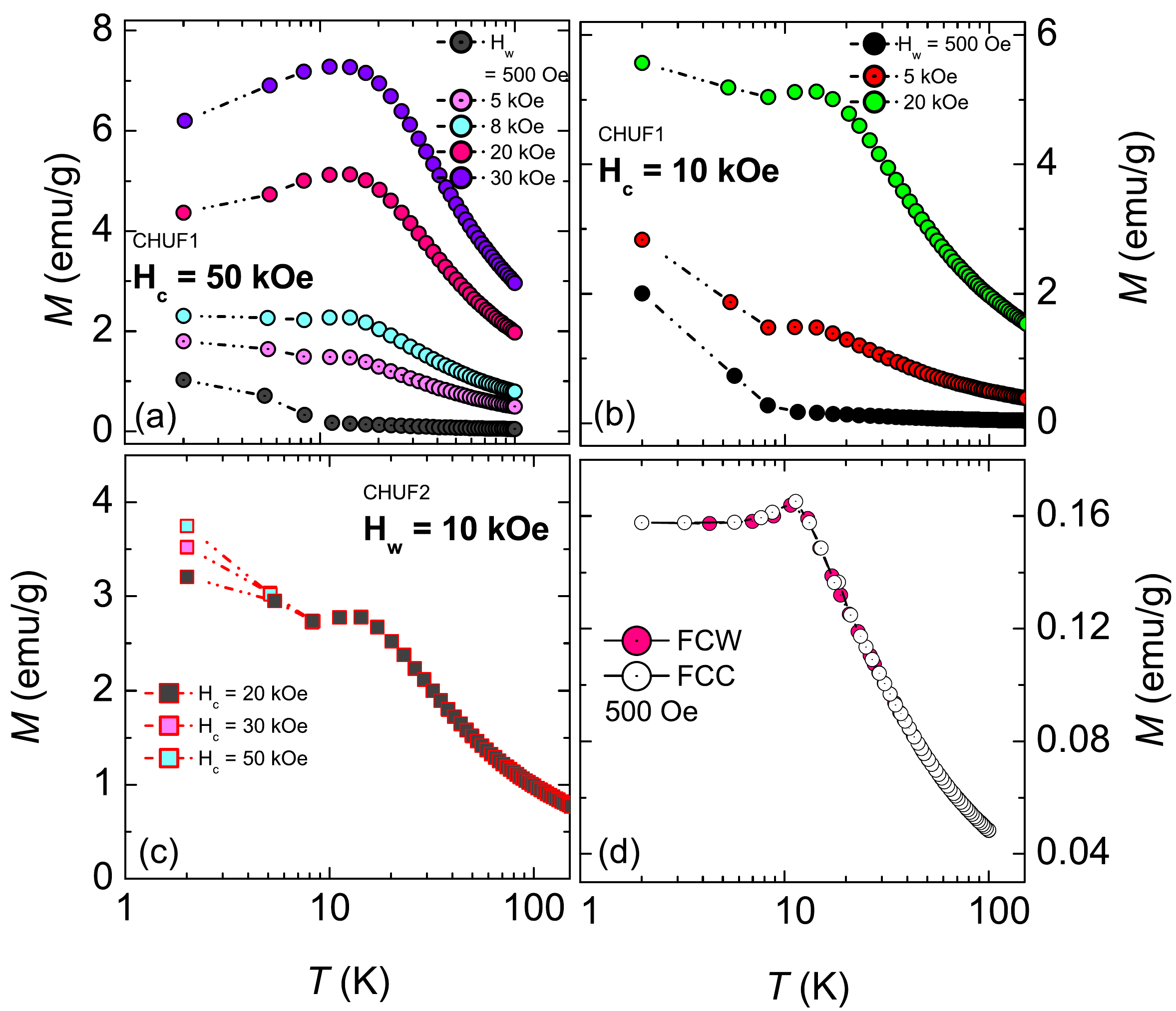}
\caption{\label{fig_chuf} (color online) (a,b) Magnetization as a function of temperature using the protocol of CHUF1: the sample was cooled in a constant field of $H_c$ = 50~kOe (a) or $H_c$ = 10~kOe (b) and the magnetization is measured while warming in various applied fields, $H_w$, as marked in the figure. (c) The magnetic response under the protocol of CHUF2: the sample was cooled in different applied fields, $H_c$, while the magnetization was measured in a constant field, $H_w$ = 10~kOe. (d) A field-cooled magnetization measurement in cooling and warming cycles do give evidence for any thermal hysteresis.}
\end{figure}
%
\indent
A previous work by MacDougall {\em et. al.,}\cite{macdougall_pnas_108_15693_2011kinetically} on single crystals of CoAl$_2$O$_4$ put forward the idea of kinetic freezing of domain walls inhibiting the long-range magnetic order. This concept was highlighted as prevailing in a broader class of frustrated magnets and, would then, offer an explanation to the reports of "anomalous" glassy behaviour. When a first order magnetic phase transition is kinetically inhibited, it can give rise to metastable states which are then referred to as "magnetic glasses". \cite{chattopadhyay_prb_72_2005kinetic} The kinetics of the phase transition can, for example, get arrested due to the coexistence of a metastable and a transformed stable state at low temperature. Several studies\cite{kushwaha_prb_80_2009low,banerjee_jpcm_21_2009conversion,roy_prb_79_2009contrasting} investigating first order magnetic phase transitions have used the "cooling-and-heating-in-unequal-fields" (CHUF) protocols to confirm metastable states at low temperatures or to distinguish the equilibrium phase from a glass-like phase. In order to gain more insight in to the spin freezing in FeAl$_2$O$_4$, we have performed CHUF measurements. The results are presented in Fig~\ref{fig_chuf}.
\\
\indent
We have performed two protocols under the CHUF measurements. In the first set (CHUF1), the sample is field-cooled from 300~K to 2~K at $H_c$ = 50~kOe (also repeated for $H_c$ = 10~kOe ). At 2~K, the field is put to zero and various applied fields of $H_w$ = 500~Oe, 5~kOe, 8~kOe, 20~kOe and 30~kOe were used to measure magnetization while warming the sample. In the second set (CHUF2), various applied fields (20, 30 and 50~kOe) were used to field-cool the sample to 2~K but the magnetization was measured in the warming cycle using a constant field of 10~kOe. The results of the first set of experiments using $H_c$ = 50~kOe (CHUF1) are presented in Fig~\ref{fig_chuf} (a) for $H_c$ = 50~kOe  and in (b) for $H_c$ = 10~kOe . A notable difference in the magnetization profile using the CHUF protocol is the increase in magnetization with increasing value of $H_w$. It is also interesting to note that the higher the value of $H_c$ used to cool through $T_g$, the more responsive is the magnetization roughly below the $T_g$. An anomaly close to the $T_a$ becomes more prominent and broadened as $H_w$ reaches 20~kOe. This could imply that some part of the kinetically arrested "glassy" phase is transformed into an equilibrium phase. The results of CHUF2 are presented in Fig~\ref{fig_chuf} (c). It can be noted that in this protocol, the magnetization remains the same irrespective of the cooling field until $T_a$ is reached. Below $T_a$, when $H_c > H_w$ there is an enhancement of magnetization. However, the temperature range measured below $T_a$ is inadequate to conclude with authority about a general trend. The result of a field-cooled magnetization measurement at 500~Oe performed in warming and cooling cycles presents no significant thermal hysteresis, see Fig~\ref{fig_chuf} (d).
\\
\indent
The purpose of our work was to search for magnetometric "signatures" of the spin liquid phase in the frustrated diamond lattice antiferromagnet exemplified by the $A$-site spinels. Frustration effects induced by the {\em n.n.n} exchange competition leads to a degenerate ground state of spin spirals in this system. The degeneracy of the ground state may be lifted by the "order by disorder" mechanism or by quenched random disorder. While the spiral spin liquid phase persists even in the presence of finite weak disorder, a spin glass or glass-like feature can develop due to the effect of disorder. Our study confirms that FeAl$_2$O$_4$ consists of a majority phase that is predominantly short-range ordered like a spin liquid and a minority phase thats remains "glassy". It is shown that the "glassy" phase transforms under the influence of suitable applied fields clearly evidenced in the CHUF experiments.
\\
\indent
We conclude that the TRM/IRM "landscapes" of  FeAl$_2$O$_4$ are quite different from those of a canonical spin glass or a superparamagnet though some similarity to DAFF systems can be assumed. Nonlinear susceptibility and critical analysis of higher harmonic terms leads to exponents very similar to those obtained for comparable systems of frustrated magnet class of materials. However, a scaling relation advised for spin glasses according to the common prevailing understanding of this class of ordering phenomena is not met. Studying the dynamics of spin relaxation using Cole-Cole formalism, it is found that a broad relaxation is present in FeAl$_2$O$_4$ as signified by a high value of $\alpha$. The "irreversible" part of dc susceptibility indicates no sign of a conventional spin glass freezing. We argue that FeAl$_2$O$_4$ resides near the spin liquid -- spin glass boundary in the $T - \eta$ phase diagram proposed for $A$-site spinels\cite{hanashima_jpsj_82_024702_2013spin}. This picture is reinforced by the results of the magnetization measurements using CHUF protocols. Our study could be extended to other spin liquid systems where, using the combination of TRM/IRM responses, Cole-Cole analysis and the CHUF protocols one could obtain "maps" of spin-liquid states in frustrated magnets.
\\
\\
%
%
HSN wishes to thank Oleg Petracic for useful discussions. HSN and RKK acknowledge FRC/URC for a postdoctoral fellowship. AMS thanks the SA-NRF (78832) and the FRC/URC of UJ for financial assistance. \\
%
%

%
\end{document}